# SBbadger: Biochemical Reaction Networks with Definable Degree Distributions

Michael A. Kochen[1], H. Steven Wiley[2], Song Feng[3], and Herbert M. Sauro[1]

[1]Department of Bioengineering, University of Washington, Seattle, WA 98105 USA, [2]Environmental Molecular Sciences Laboratory, Pacific Northwest National Laboratory, Richland, WA 99354, USA, [3]Biological Science Division, Pacific Northwest National Laboratory, Richland, WA 99354, USA

## Abstract

**Motivation:** An essential step in developing computational tools for the inference, optimization, and simulation of biochemical reaction networks is gauging tool performance against earlier efforts using an appropriate set of benchmarks. General strategies for the assembly of benchmark models include collection from the literature, creation via subnetwork extraction and de novo generation. However, with respect to biochemical reaction networks, these approaches and their associated tools are either poorly suited to generate models that reflect the wide range of properties found in natural biochemical networks or to do so in numbers that enable rigorous statistical analysis.

**Results:** In this work we present SBbadger, a python-based software tool for the generation of synthetic biochemical reaction or metabolic networks with user-defined degree distributions, multiple available kinetic formalisms, and a host of other definable properties. SBbadger thus enables the creation of benchmark model sets that reflect properties of biological systems and generate the kinetics and model structures typically targeted by computational analysis and inference software. Here we detail the computational and algorithmic workflow of SBbadger, demonstrate its performance under various settings, provide sample outputs, and compare it to currently available biochemical reaction network generation software.

**Availability and Implementation:** SBbadger is implemented in Python and is freely available at https://github.com/sys-bio/SBbadger and via PyPi at https://pypi.org/project/SBbadger/. Documentation can be found at https://SBbadger.readthedocs.io.

# Introduction

Computational simulation and analysis of biochemical reaction networks is a valuable complement to experimentation as it enables the generation of hypotheses regarding the governance of system dynamics and predictions of treatment responses that can suggest productive experiments (Aldridge *et al.*, 2006; Le Novère, 2015). Such analysis depends on a multi-step workflow that incorporates multiple computational tools, all of which can influence the reliability and feasibility of the resulting hypotheses and predictions. This workflow begins with the construction of a biochemical reaction topology, a task traditionally done with expert knowledge, extensive search of the literature, and additional experimentation. When uncertainties in network structure are apparent, methods for selecting a best candidate model or an ensemble of models that best explain available data can be utilized. Many variants of information theoretic, Bayesian, and other model selection approaches may be used here, as well as numerous tools to carry them out (Burnham and Anderson, 1998; Kirk *et al.*, 2013). Once a reaction network or ensemble of networks are constructed, rate laws must be imposed on those reactions and the associated parameters must be calibrated to experimental data. Here again, numerous existing approaches and tools are available (Vrugt, 2016; Thomas *et al.*, 2016; Shockley *et al.*, 2018; Ashyraliyev *et al.*, 2009; Raue *et al.*, 2013). Recently, there has been increased interest in data-driven approaches that simultaneously infer the reaction topology and calibrated parameters. One such approach, sparse identification of nonlinear dynamics (SINDy), employs regularized regression to enforce sparsity on a super-set of potential reactions (Mangan *et al.*, 2016; Hoffmann *et al.*, 2019; Burnham *et al.*, 2008). Bayesian methods take a similar approach but use Monte Carlo sampling while enforcing sparsity with an appropriate prior (Pan *et al.*, 2012; Jiang *et al.*, 2022; Galagali and Marzouk, 2015). Neural networks have also been used to infer the reaction topologies and in some cases parameters and can be cast in a form that is directly interpretable as a reaction network (Ji and Deng, 2021). The final step in the workflow is simulation of the model. This can be done via numerical integration of a representative system of ordinary differential equations (ODE), stochastically using a stochastic simulation algorithm (SSA), or some other form of simulation. Once again there are many algorithms and tools to carry this out (Postawa *et al.*, 2020; Hoops *et al.*, 2006; Somogyi *et al.*, 2015; Gupta and Mendes, 2018; Materi and Wishart, 2007; Choi *et al.*, 2018). New methods and/or tools are regularly introduced for each of these steps and comparative benchmarking is essential for assessing improvements in algorithmic speed and accuracy over existing methods and software and to determine the most appropriate approach for a given problem.

Benchmarking, of course, requires benchmark models to compare against. A common approach to obtain benchmark models of reaction networks is to assemble a collection of existing models from the literature, or from model repositories like BioModels (Malik-Sheriff *et al.*, 2020) or JWS Online (Olivier and Snoep, 2004). The aim is to collect a diverse set of models with a wide range of species and parameter counts, different kinetic formalisms, and of various types (gene regulatory networks, signaling pathways, metabolic, etc.). An obvious advantage to this approach is that these models are already characteristic of models that computational tools are typically applied to. Another potential advantage is that experimental data may accompany such models. However, concomitant data may be sparse or nonexistent, thus requiring supplementation with synthetic data. There is also a lack of breadth and depth of available models. Relevant variations in model properties may be extensive and obtaining a sufficient number of benchmarks for statistical comparisons may not be possible. Recent efforts at collecting benchmarks used from 6 to 142 models (Villaverde *et al.*, 2019; Städter *et al.*, 2021; Hass *et al.*, 2019; Gennemark and Wedelin, 2009; Villaverde *et al.*, 2015).

Other approaches for benchmark generation include subnetwork extraction and de novo model generation. Subnetwork extraction is often used to produce benchmark gene regulatory networks (Schaffter *et al.*, 2011; Li *et al.*, 2009; Van den Bulcke *et al.*, 2006). The subnetworks are extracted from existing GRNs, with various strategies, and assigned kinetic equations. This approach increases the likelihood of preserving the graphical properties of the source network and produces GRNs that are more similar to biological GRNs than those produced from standard network generation models like random (Erdős and Rényi, 1960), small-world (Watts and Strogatz, 1998), and scale-free (Albert and Barabási, 2000). Unfortunately, this approach is impractical for biochemical reaction networks because effective subnetwork extraction requires the large network size and simplistic structure that characterize GRNs.

De novo generation of synthetic biochemical reaction networks is, to our knowledge, currently limited to SMGen as recently described in (Riva *et al.*, 2022). SMGen creates fully connected models with definable species and reaction counts, definable reactant and product maximums per reaction, and multiple parameter distribution options. However, these models are limited to mass action kinetics and there is no mechanism to control network properties like outdegree and indegree distributions. Ideally, benchmark models should resemble the complete range of typical models, which requires a high degree of flexibility in model creation.

In this work we present SBbadger, a python-based generator of synthetic biochemical reaction and metabolic models with customizable directed degree distributions, multiple types of available kinetic formalisms, four different parameter sampling distributions and many other options to produce tailored benchmark models. SBbadger should thus be particularly useful in the development of new modeling and network analysis tools.

## Methods and Algorithms

### Overview

SBbadger constructs synthetic biochemical reaction networks guided by user specifications. Model construction is split into three modules that 1) generate degree frequency distributions, 2) generate reaction topologies that conform to the degree frequencies, and 3) impose rate laws on the networks (Figure 1).

The primary method for producing an ensemble of such models is the parallelized `generate.models()` method (Figures 1-2). A serialized version, `generate_serial.models()`, is also available, which is true for all methods described below. Another option is the `generate.model()` method which can be used at the command line to return a single model as an Antimony string, a human readable text based model format designed for easy model creation and modification (Smith *et al.*, 2009).

In addition to the models, files containing information regarding the distributions and reaction networks are also created (Figure 1). This allows for new networks to be created from previously created degree distributions with the `generate.networks()` method or the imposition of different rate laws on previously generated networks with `generate.rate_laws()`. The `generate.distributions()` method can be used to construct stand-alone degree frequency distributions for use with the `generate.networks()` and `generate.rate_laws()` methods or with other modeling software applications. Three special network types are also available using the `generate.linear()`, `generate.cyclic()`, and `generate.branched()` methods. The use of all these methods and the available arguments to them are further detailed below, in the supplementary material, and in the documentation at https://SBbadger.readthedocs.io.

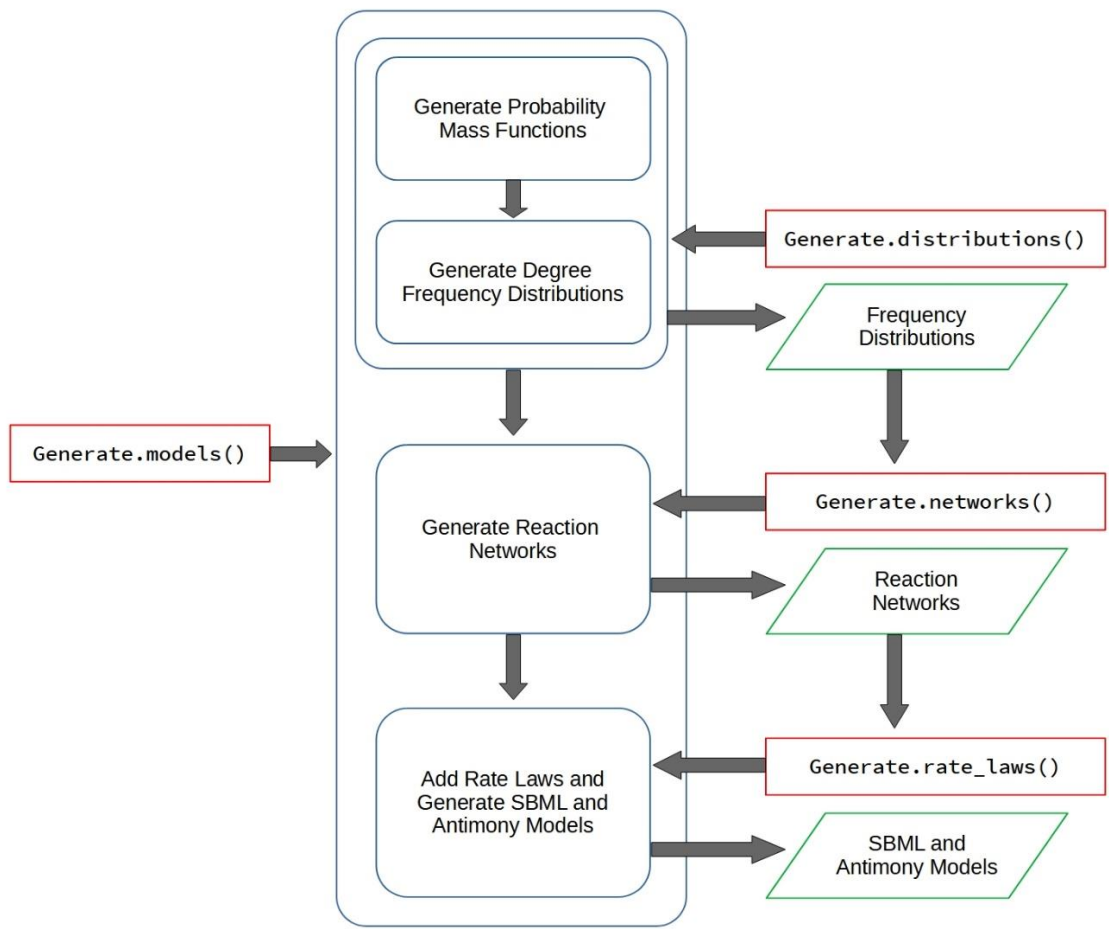

**Figure 1.** SBbadger workflow. Red rectangles: methods, blue rounded rectangles: processes, green rhombuses: output files. The stages of model creation can be run together or individually. The `generate.models()` method, and its variants, generates frequency distributions, reaction networks, and kinetic rate laws sequentially to produce a set of models in Antimony and SBML formats. The `generate.distributions()` method produces only the frequency distributions which can then be used as input to the `generate.networks()` method to produce reaction networks. The reaction networks can, in turn, be used as input for the `generate.rate_laws()` method to produce a final set of models.

```python
from SBbadger import generate
from scipy.special import zeta
from scipy.stats import zipf

def in_dist(k):
    return k ** (-2) / zeta(2)

def out_dist(k):
    return zipf.pmf(k, 3)

if __name__ == "__main__":

    generate.models(
        group_name='example',
        n_models=1,
        n_species=50,
        out_dist=out_dist,
        in_dist=in_dist,
        out_range=[1, 10],
        in_range=[1, 10],
        rxn_prob=[.35, .3, .3, .05],
        kinetics=['mass_action', ['loguniform', 'loguniform', 'loguniform'],
                  ['kf', 'kr', 'kc'],
                  [[0.01, 100], [0.01, 100], [0.01, 100]]],
        ic_params=['uniform', 0, 10],
        dist_plots=True
    )
```

**Figure 2.** Example of the `generate.models()` method. `group_name`: output directory, `n_models`: number of generated models, `n_species`: number of species per model, `out_dist`/`in_dist`: the outdegree and indegree functions from which the distributions and degree frequencies are derived, `out_range`/`in_range`: the initial degree ranges for the outdegree and indegree distributions, `rxn_prob`: the probability distribution for selection from the four available reaction patterns, `kinetics`: defines the type of kinetics and the distributions of the associated parameters, `ic_params`: defines the distribution of initial values.

## Degree Distributions

The outdegree and indegree of a node in a directed network are the number of edges extending from, and incident on, that node respectively (in this work network nodes and model species are equivalent). Degree frequencies are obtained by sampling from distributions that are derived from user-provided functions. These functions can be defined explicitly or used as function wrappers around existing functions like, for example, the Scipy probability mass function `zipf.pmf` in Figure 2. The argument to these functions, $k$, is interpreted as an integer degree value and the return value for each $k$ is the relative weight for that degree in the resulting discrete distributions. For example, the `in_dist()` function in Figure 2 is an explicit representation of the zeta (power-law) distribution with an $\alpha$ value of 2. Note that normalizing constants, like $zeta(\alpha)$, are not strictly necessary; SBbadger will automatically normalize these functions for reasons described below. This also permits continuous distributions, or general functions, to be defined and reinterpreted as discrete distributions. The `out_dist()` function in Figure 2 is the zeta distribution as defined in Scipy with an $\alpha$ value of 3. The distribution functions are themselves arguments to the model or degree distribution generating methods (Figure 2).

The renormalization of the degree distributions is done to account for their necessary truncation. Sampling node degrees from unbounded distributions runs the risk of producing networks with nodes that have hundreds or thousands of edges, leading to model species that take part in as many reactions. This unrealistic scenario becomes more likely as the number of nodes per model is increased. Therefore, a minimum expected value is placed on the degree frequencies. For example, if the minimum expected frequency is 1 (default), then every degree in the distribution must have an expected value of at least 1 node when the specified number of nodes (`n_species`, Figure 2) are allocated over the distribution. The threshold can be set via the `min_freq` argument.

Truncation and renormalization are done iteratively. One of two algorithms is deployed depending on whether a degree range (Examples: `out_range/in_range`, Figure 2) has been specified. If no range is provided, the algorithm begins with a distribution consisting of 1-degree nodes (Algorithm 1). The distribution is iteratively expanded with the probabilities for each degree updated based on relative values of the output from the user-provided functions. Expected frequency values are then updated based on the new probabilities and the number of nodes in the network. This process continues until an expected frequency is found that falls below the threshold (Table S1, Figure S1). The broadest probability distribution that does conform to the threshold is then selected.

**Algorithm 1:** Discrete Distribution and Expected Degree Frequencies for Unbounded Distribution Functions
**Inputs:**
        Expected Frequency Threshold (EFT)
        Distribution Function
**Initialization:**
        Set the degree range as [$D_{initial}$, $D_{final}$] = [1, 1]
        Set the probability mass function (PMF) to 1 for degree 1
        Set the expected degree frequencies (EDF) to `n_species` for degree 1

**While** EDF(j) > EFT for each degree j in the EDF
        Set cPMF (current PMF) = PMF
        Set cEDF (current EDF) = EDF
        $D_{final}$ = $D_{final}$ + 1
        Recalculate PMF
        Recalculate EDF

**Return** cPMF

With algorithm 1 we assume that the provided distribution function is heavily skewed to the right, such as with a power-law distribution. If this assumption fails, then the expected frequencies for lower degree values will fall below the threshold before the higher degree values and SBbadger will return an error that the function is invalid. The solution is to provide a degree range. Then Algorithm 2 will iteratively remove the degree with the lowest frequency, renormalize the distribution, and recalculate the frequencies until all frequencies in the distribution are above the required minimum (Table S2, Figure S2).

**Algorithm 2:** Discrete Distribution and Expected Degree Frequencies for Bounded Distribution Functions
**Inputs:**
        Expected Frequency Threshold (EFT)
        Distribution Function
        Degree Range [$D_i$, $D_{i+n}$]
**Initialization:**
        Expand the degree range: [$D_i$, $D_{i+1}$, $D_{i+2}$, … $D_{i+n}$]
        Calculate probability mass function (PMF) for each j in [i, i+n]
        Calculate the expected degree frequencies (EDF) for each j in [i, i+n]

**While** there exists a degree j such that EDF(j) < EFT
        Remove $D_j$ for $EDF_j$ = Min(EDF)
        Recalculate PMF
        Recalculate EDF

**Return** PMF

In the absence of one or both distributions, nodes for those edges will be selected at random. However, if functions for both the outdegree and indegree distributions are provided and are different, the computed distributions must be reconciled so that the total number of expected indegree and outdegree edges are similar (Algorithm 3). This is done by iteratively reducing the number of nodes in the distribution with the higher expected edge count and adjusting the bounds of the distribution accordingly (Table S3). Once the number of expected edges for each distribution are as close as possible the algorithm terminates. Because the final frequency distributions are sampled from these probability distributions, and the number of out-edges and in-edges must be the same, multiple resamplings may be needed before obtaining compatible distributions. The untouched distribution is first sampled up to the number of species in the network. The trimmed distribution is then sampled up to the number of edges in the first distribution. It is possible that the distributions cannot be made close enough to get matching edge counts. In that case, the algorithm will terminate after a set number of attempts and return an error. If successful, a consequence of having different outdegree and indegree distributions is that some nodes will have no outgoing or incoming edges.

---

**Algorithm 3:** Outdegree and Indegree Distribution/Degree Frequency Reconciliation
**Inputs:**
        Outdegree/Indegree probability mass functions (PMF)
        Number of species `n_species`
**Initialization:**
        Calculate Outdegree/Indegree expected degree frequencies (EDF)
        Calculate total expected out-edge and in-edge counts.
        Choose Min(out-edges, in-edges) as edge number target (ENT)
        Choose Distribution with Max(out-edges, in-edges) as PMF/EDF to Modify mPMF/mEDF
        Set modifiable number of species mSP = `n_species`

**While** Sum(mEDF) > ENT
        Set cPMF (current PMF) = mPMF
        Set cEDF (current EDF) = mEDF
        mSP = mSP - 1
        Recalculate mPMF
        Recalculate mEDF

**If** ABS(ENT - Sum(cEDF)) < ABS(ENT - Sum(mEDF))
        **Return** cPMF
ELSE
        **Return** mPMF

Optional frequency distribution figures can be output with `dist_plots=True` (Figures 2, 3A). In a run of $n$ models with defined distribution functions each model will have its own sampling of the accepted probability distribution. Species (nodes) are then randomly assigned to the degree frequencies (Figure 3B) and will have edges to match by the end of network construction.

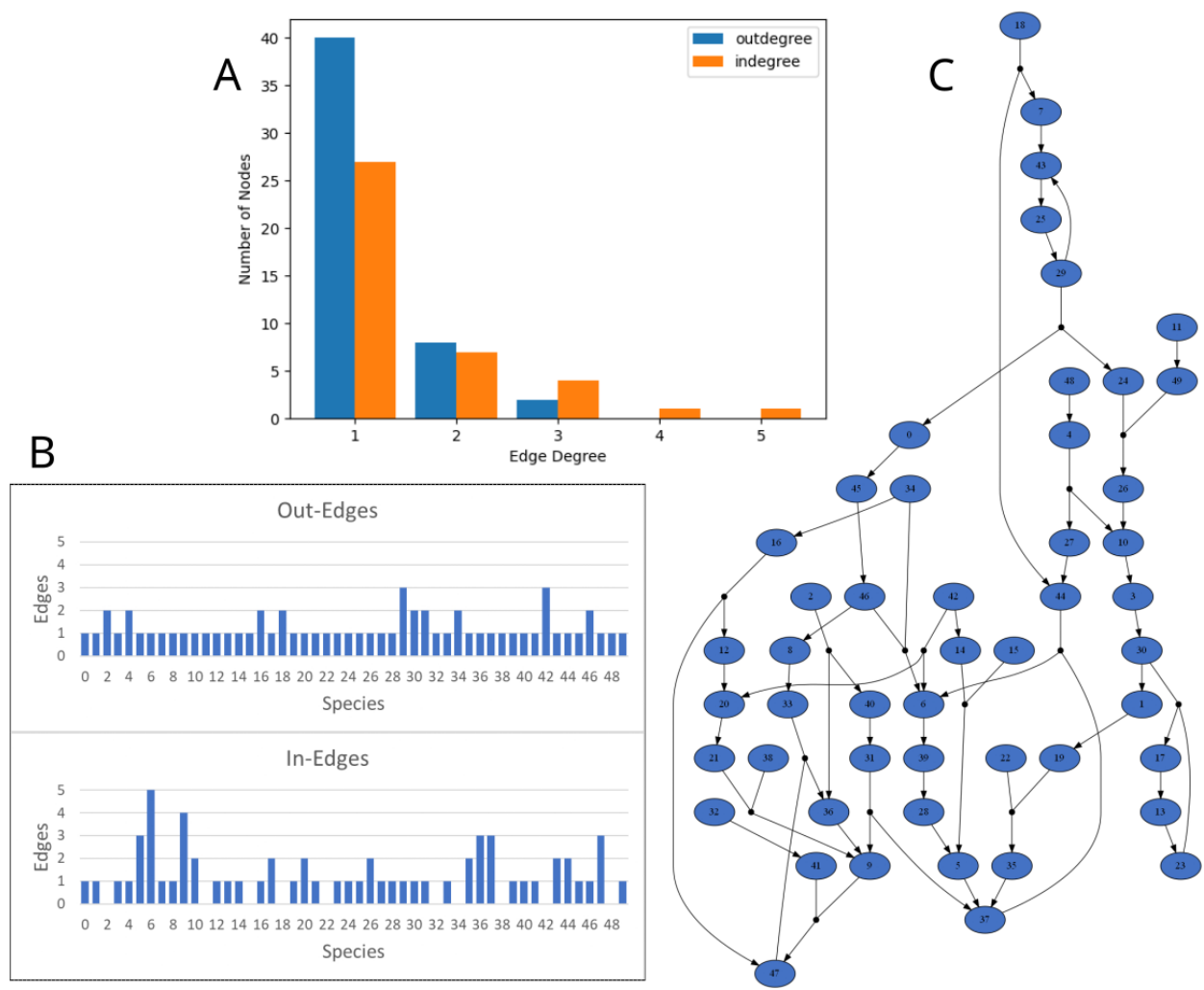

**Figure 3.** Example outdegree and indegree frequency distributions for 50 nodes from power law functions with $\alpha$ values of 3 and 2 for the outdegree and indegree functions respectively. (A) Sampled frequency distributions figure produced by SBbadger. Total number of edges for both the indegree and outdegree distributions is 62. (B) Species are randomly assigned to the distributions to obtain the edge frequencies per node. As expected, some nodes lack any in-edges as that distribution required downsizing. (C) Depiction of a reaction network generated from the edge frequencies in (B).

Joint distributions are allowed but must be largely symmetrical to comply with the in-edge/out-edge constraint. For example, a bivariate normal distribution might be defined and invoked as in Figure S3. The range for the joint distribution is symmetrical. The algorithm to truncate and normalize the joint probability distribution is similar to the 1-dimensional case but works on a 2-dimensional grid of probabilities. With separately defined outdegree and indegree distribution functions, each will have their own (reconciled) probability distributions that are separately sampled. In the joint distribution case, there is one 2-dimensional distribution that the outdegree and indegree frequency distributions are sampled from simultaneously. The algorithm will return an error if it fails to equate the out-edges and in-edges after a set number of sampling attempts. For added flexibility, lists of predefined probability distributions or degree frequencies can be supplied instead (Text S1).

**Reaction Networks**

Construction of fully connected networks proceeds iteratively by probabilistically selecting both a reaction pattern and the species to serve as reactants and products at each iteration. Each added reaction creates new out-edges and in-edges for the reactants and products respectively. Network construction terminates when the number of out-edges and in-edges for each species in the network match their designated outdegree and indegree frequencies. There are currently four supported reaction patterns (Figure 4A). The probability of selecting a given pattern is defined in the `rxn_prob` argument which takes a list of probabilities in the order [Uni-Uni, Bi-Uni, Uni-Bi, Bi-Bi] (Figure 2). Once a reaction pattern is selected, reactants and products are chosen probabilistically based on each species' current number of unconnected edges relative to the total number of unconnected edges. For example, if the selected reaction pattern is Uni-Uni, then the probability of selecting species 0 from the example in Figure 3B as a reactant is 1/62 because that species has 1 of the 62 total available out-edges. However, if the reaction pattern were Uni-Bi then only those reactants with at least two available out-edges are eligible to be selected (species 2, 4, 16, 18, 29, 30, 34, 42, and 46 in Figure 3B). If no such species are available, then a new reaction pattern is chosen. The available edges for all species selected to take part in the reaction are updated for the next iteration.

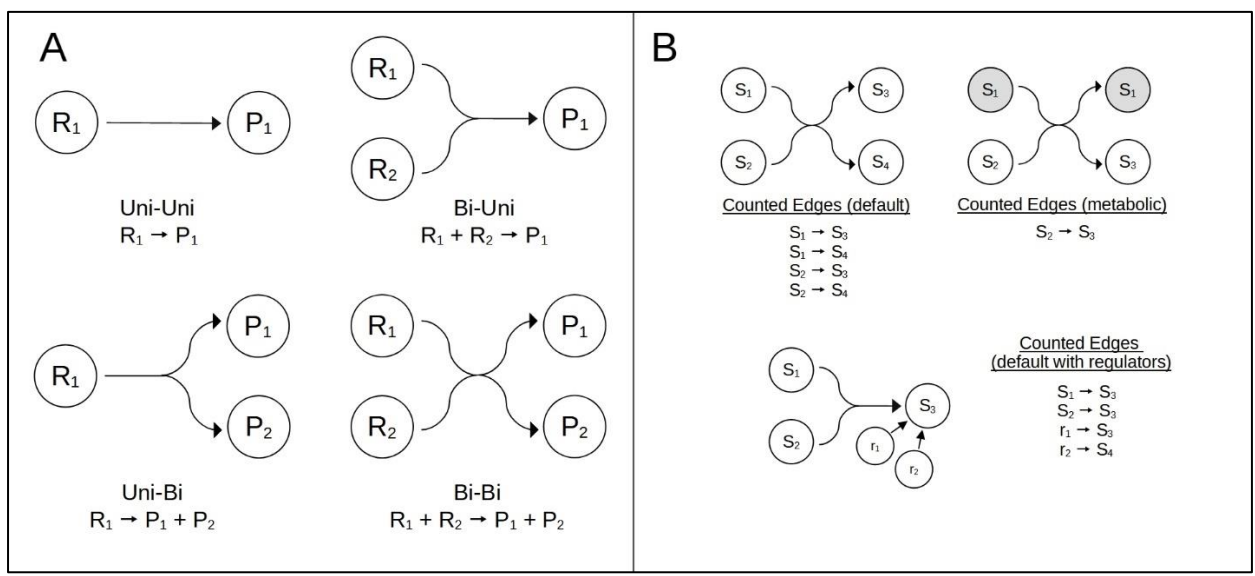

**Figure 4.** Network construction. At each iteration a reaction pattern is probabilistically selected from the set of four in (A). The reaction pattern probability distribution is user-definable (Figure 2). (B) Once the reaction pattern is selected, reactants, products, and possibly regulators are selected probabilistically based on the availability of edges for each species. By default, a new edge is designated from each reactant or regulator to each product. In the "metabolic" edge scheme, only edges between consumed and produced species are counted. The available out-edges and in-edges for the selected species are then updated for the next iteration and the algorithm terminates when all available edges are exhausted.

There are two schemes in which reaction edges are counted against the available in-edges and/or out-edges for the species chosen as reactants and/or products respectively (Figure 4B). The default scheme simply counts an edge from each reactant to each product. The second counts only those edges that represent the flow of metabolites through the network. For example, if a species is chosen as both a reactant and a product, as for $S_1$ in the reaction $S_1 + S_2 \rightarrow S_1 + S_3$, then $S_1$ is implicitly considered an enzyme or modifier and only the edge from $S_2$ to $S_3$ is counted against their available out-edges and in-edges respectively. The argument to enforce metabolic edges is `edge_type='metabolic'`. If no distribution is defined for either the outdegree or indegree distributions, then the reactants or products respectively are chosen at random. Of the available kinetic rate laws (next section) there are three (modular, generalized mass action, saturable and cooperative; see next section for details) that allow for up to three additional regulatory species. In the default edge counting scheme these regulators are treated as reactants while in the metabolic scheme these interactions are not counted against available edges.

Several other options are also available. Reversible reactions can be included with the argument `rev_prob` which takes a value from 0 to 1 as the probability of declaring a reversible reaction at each iteration. However, the reverse edges associated with them do not count against the available edges. A filter for mass balance violating reactions is available with the `mass_violating_reactions=False` argument, ruling out reactions such as $S_1 + S_2 \rightarrow S_1$. Mass balance can be ensured at the network level with the `mass_balanced=True` argument. This option uses linear programming to check for inconsistencies in the stoichiometry matrix at every iteration of network construction (Shin and Hellerstein, 2021). Network figures can be optionally produces with the `net_plots=True` argument. SBbadger makes use of the Pydot python package (https://github.com/pydot/pydot) and can use any of the Pydot layouts through the `net_layout` argument. Figure 3C provides an example with `net_layout='dot'` (default).

**Rate Laws**

After completion of the reaction networks, kinetic rate laws are imposed on each reaction therein. Six kinetic formalisms are available: mass action, generalized mass action (Alves *et al.*, 2008), lin-log (Sauro, 2011; Visser and Heijnen, 2003), generalized Michaelis-Menten (Hanekom, 2006), saturable and cooperative (Sorribas *et al.*, 2007), and modular (Liebermeister *et al.*, 2010), with each having their own set of parameters. The mathematical details for each are outlined here with definable kinetic parameters in <span style="color:red">red</span>. Kinetic parameters, as well as initial conditions (`ic_params`, Figure 2), are chosen randomly from user-defined ranges from four possible distributions: uniform, log-uniform, normal, and log-normal. In general, the argument to define the kinetics is

`kinetics=[type, [dist₁, dist₂, …], [k₁, k₂, …], [d₁₁, d₁₂], [d₂₁, d₂₂], …]`

where `type` is the type of kinetics, `distᵢ` is the distribution type for the $ith$ parameter, `kᵢ` is the $ith$ parameter, and `dᵢⱼ` is the $jth$ distribution parameter for the $ith$ kinetic parameter. Computational examples for each of the kinetic types below can be found in the Supplementary material. A degradation rate parameter <span style="color:red">*deg*</span> can be optionally added to each type of rate laws and will be applied to each species if present.

*Mass action.* Mass action rate laws take the form

$$kf * \prod_i R_i - kr * \prod_i P_i$$

for reversible reactions or

$$kc * \prod_i R_i$$

for irreversible reactions (Figure 2). $kf$, $kr$, and $kc$ are the forward, reverse, and catalytic rate parameters and $R_i$ and $P_i$ are reactants and products respectively. An additional option for mass action kinetics is to define parameter distributions for individual reaction patterns, expanding the number to 12 (figure S4).

*Generalized mass action.* Generalized mass action rate laws take the form

$$\left(kf \prod_i R_i^{ko_{R_i}} - kr \prod_i P_i^{ko_{P_i}}\right) * \prod_i Rg_i^{kor_{Rg_i}}$$

for reversible reactions or

$$kc \prod_i R_i^{ko_{R_i}} * \prod_i Rg_i^{kor_{Rg_i}}$$

for irreversible reactions. This version of mass action kinetics allows for regulators ($Rg$) and kinetic orders for the reactants and products ($ko$), as well as the regulators ($kor$). The kinetic orders for the regulators are allowed to have negative values to impart inhibitory effects (figure S5).

*Lin-Log.* Lin-Log rate laws take the form

$$v * \left(1 + \sum_i \left(T_i^S * \log \frac{S_i}{rc_{S_i}}\right) - \sum_i \left(T_i^P * \log \frac{P_i}{rc_{P_i}}\right)\right)$$

where $v$ is the reference rate, $S_i$ and $P_i$ are the reactants and products respectively, $rc_{S_i}$ and $rc_{P_i}$ are their respective reference concentrations and $T_i^S$ and $T_i^P$ are their respective stoichiometric coefficients (Figure S6).

***Generalized Michaelis-Menten (Hanekom).*** The Generalized Michaelis-Menten rate laws are dependent on the type of reaction and take the forms:

- UNI − UNI: $\quad v * \alpha * \dfrac{1-\Gamma/k_{eq}}{1+\alpha+\pi}$
- BI − UNI: $\quad v * \alpha_1 * \alpha_2 * \dfrac{1-\Gamma/k_{eq}}{1+\alpha_1+\alpha_2+\alpha_1\alpha_2+\pi}$
- UNI − BI: $\quad v * \alpha * \dfrac{1-\Gamma/k_{eq}}{1+\alpha+\pi_1+\pi_2+\pi_1\pi_2}$
- BI − BI: $\quad v * \alpha_1 * \alpha_2 * \dfrac{1-\Gamma/k_{eq}}{(1+\alpha_1+\pi_1)(1+\alpha_2+\pi_2)}$

where $v$ is the maximum reaction velocity, $\alpha_i = S_i/k_{S_i}$ where $S_i$ are reactants and $k_{S_i}$ is the half-saturation constants for $S_i$, $\pi_i = P_i/k_{P_i}$ where $P_i$ are products and $k_{P_i}$ is the half-saturation constants for $P_i$, $\Gamma = \prod_i P_i / \prod_i S_i$ is the mass action ratio and $k_{eq}$ is the equilibrium constant (Figure S7).

***Saturable and Cooperative.*** Saturable and cooperative kinetics are based on a Taylor series approximation around an operating point. The rate laws take the form

$$v \frac{\left(\prod_i R_i^{n_{R_i}} - \prod_i P_i^{n_{P_i}}\right) * \prod_i Rg_i^{nr_{Rg_i}}}{\left(\prod_i\left(k_{R_i} + R_i^{n_{R_i}}\right)\prod_i\left(k_{P_i} + P_i^{n_{P_i}}\right)\prod_i\left(k_{Rg_i} + Rg_i^{n_{Rg_i}}\right)\right)}.$$

The parameters v, n, nr, and k are derivative of other constants, details of which can be found in (Sorribas *et al.*, 2007). The kinetic order for the regulators, $nr$, is distinct from those of the reactants and products, $n$, because it is allowed to be negative for inhibitory effect (Figure S8).

***Modular (Liebermeister).*** Modular rate laws are a special case in that there are multiple components and five types of rate-laws (Figure S9). The rate laws take the general form

$$F * \frac{T}{D + D_{reg}}.$$

The following describes each of these components and the different types of rate laws.

***T*** is mass action like and takes the form

$$k_f \prod_i \alpha_i^{m_{S_i}} - k_r \prod_i \pi_i^{m_{P_i}}$$

where $k_f$ and $k_r$ are the forward and reverse rates respectively, $\alpha_i = S_i/k_{S_i}$ and $\pi_i = P_i/k_{P_i}$ where $S_i$ and $P_i$ are reactants and products respectively, $k_{S_i}$ and $k_{P_i}$ are reactant and product constants, and $m_{S_i}$ and $m_{P_i}$ are reactant and product turnover rates.

**D** represents the five possible rate laws with the following forms:

- CM: $\prod_i (1 + \alpha_i)^{m_{S_i}} + \prod_i (1 + \pi_j)^{m_{P_i}} - 1$
- DM: $1 + \prod_i \alpha_i^{m_{S_i}} + \prod_i \pi_i^{m_{P_i}}$
- SM: $\prod_i (1 + \alpha_i)^{m_{S_i}} * \prod_i (1 + \pi_i)^{m_{P_i}}$
- FM: $\sqrt{\prod_i \alpha_i^{m_{S_i}} * \prod_i \pi_i^{m_{P_i}}}$
- PM: $1$

**F** represents allosteric regulation and takes the form:

$$\prod_i (\rho_i^A + [1 - \rho_i^A]\beta_i)^{m_{a,A_i}} \prod_i (\rho_i^I + [1 - \rho_i^I]\gamma_i)^{m_{a,I_i}}$$

where $\rho_i^A$ and $\rho_i^I$ are relative basal rates for the activator and inhibitors respectively and range between 0 and 1, $p_i$ and $q_i$ are activator and inhibitor regulation numbers

$$\beta_i = \frac{A_i/k_{A_i}}{1 + A_i/k_{A_i}}$$

where $A_i$ is an activator and $k_{R_i^A}$ is its associated constant, and

$$\gamma_i = \frac{1}{1 + I_i/k_{I_i}}$$

where $I_i$ is an inhibitor and $k_{R_i^I}$ is its associated constant.

**$D_{reg}$** is the specific regulation term and takes the form

$$\sum_i \left(k_{A_i}/A_i\right)^{m_{d,A_i}} + \sum_i \left(I_i/k_{I_i}\right)^{m_{d,I_i}}$$

where the constants mirror those of the allosteric case.

## Results and Discussion

Getting started with SBbadger is easily done at the command line. The following example will return a randomly generated model as an Antimony string with 10 species and mass action kinetics (both default properties; see Figure 2 for default parameter distributions).

```
from SBbadger import generate
model = generate.model()
```

Ensembles of models with user-defined properties can just as easily be created. The following example will create 100 random models with 50 species each and mass action kinetics. By default, they are deposited in a `models/test/` directory.

```
from SBbadger import generate
generate.models(n_models=100, n_species=50)
```

SBbadger generated models have species designated as either boundary (`ext`) or floating (`var`). Boundary species are those with only out-edges or in-edges and will be given a fixed value when simulated via Tellurium or Roadrunner (Choi *et al.*, 2018; Somogyi *et al.*, 2015). This provides more realistic signaling and metabolic scenarios than depletion of inputs to the system. It also increases the likelihood of obtaining a diverse array of dynamics, from simple systems that quickly attain non-zero steady states, to complex systems with multiple levels of oscillatory behavior (Figure 5).

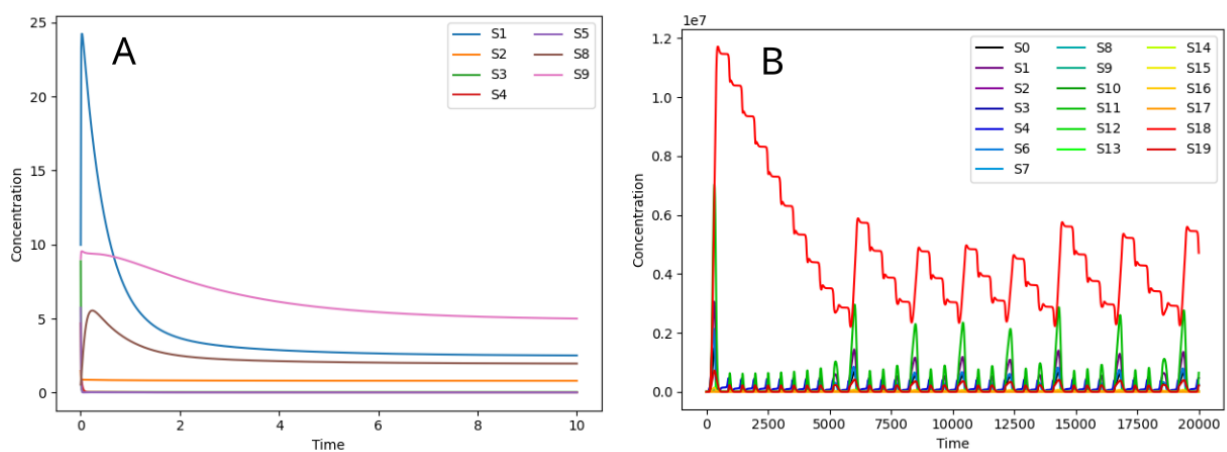

**Figure 5.** Example dynamics for models generated by SBbadger. (A) A 10-species system with three fixed boundary species (S0, S6, and S7). All floating species quickly obtain a steady state. (B) A 20-species system with one boundary species (S5) and complex oscillatory behavior.

SBbadger uses python's multiprocessing capabilities to speed up model generation. Runtimes for the serial and parallel generation of 1000 models for three distribution setups and of various sizes are given in Figures S10 and 6 respectively. These runs were done using 1/8 cores of an 8-core Ryzen-7 4700U processor. The number of cores to be used can be set with the `n_cpus` argument. Parallel model generation was at least 5x faster (Table S4).

Time to completion and the number of reactions in a generated network are highly dependent on the number of species and the degree distributions (Figure 6A-B). When distributions are defined, reaction nodes are chosen from the set of nodes with unmatched edges. This constrains the number of reactions needed to complete the network. With no defined distributions, species are randomly chosen at each iteration until every species participates in at least one reaction. This results in a slower completion time and higher number of reactions. For example, the 800 species run in Figure 6 has an average of ~2156 reactions and completion time of ~2911s. In the case with both an indegree and outdegree distribution, (power law exponents of 2 and 3 respectively), the distribution with a smaller expected value of reactions, here the outdegree distribution, sets the constraint and the other distribution is adjusted to match (see Degree Distributions above). Thus, the two-distribution example has a lower average number of reactions (~708 vs. ~1085) and lower runtime (~731 vs. ~1065) than the example with a single distribution and power law exponent of 2.

SBbadger and SMGen demonstrate comparable model generation runtimes (Riva *et al.*, 2022). Figure 7 displays the SBbadger 400 species 1000 model runs from Figure 6, and three comparable runs with SMGen. While the number of reactions generated with SBbadger is dictated by the number of species and the degree and reaction distributions (a minimum *can* be set when no distributions are defined), the number of reactions generated by SMGen is set by the user. The three SMGen runs had 101 (the minimum allowed with 400 species and max reactants/products of 2/2 per reaction), 400 (parity with the species number), and 984 (the average number of reactions generated for the SBbadger random run) reactions. The number of reactions has a significant impact on the runtime for SMGen, which overlaps those of SBbadger. However, with either software, model generation time is unlikely to be a greater bottleneck than analysis, as tasks such as model calibration can easily take orders of magnitude more time. The choice of software would thus come down to the feature sets. The major differences between the two platforms are described in Table 1.

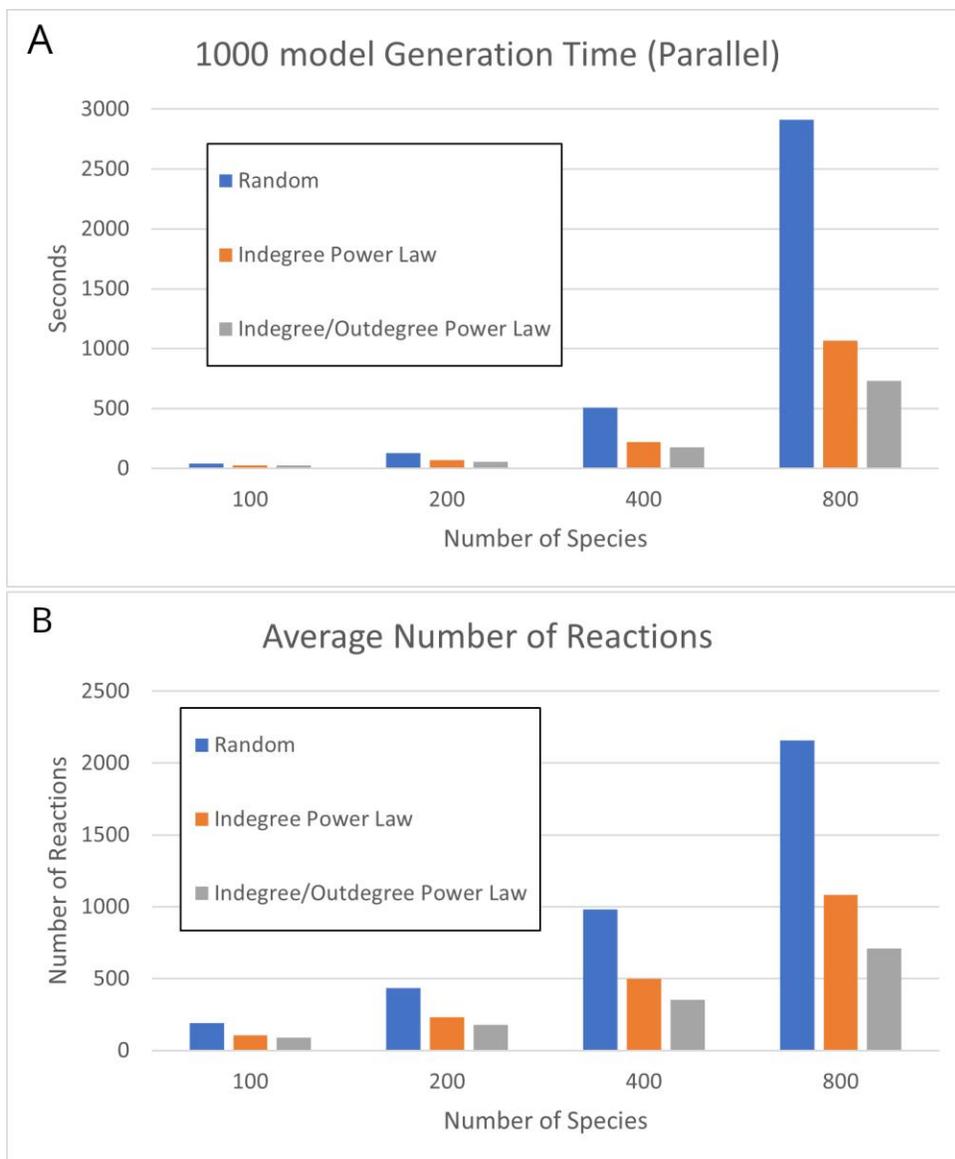

**Figure 6.** 1000 generated models with increasing numbers of species (100, 200, 400, and 800) and three different degree distributions: Random, power law indegree distribution with α=2, and power law indegree/outdegree distributions with $\alpha = 2$, and $\alpha = 3$ respectively. (A) Time to completion. (B) Average number of reactions per model.

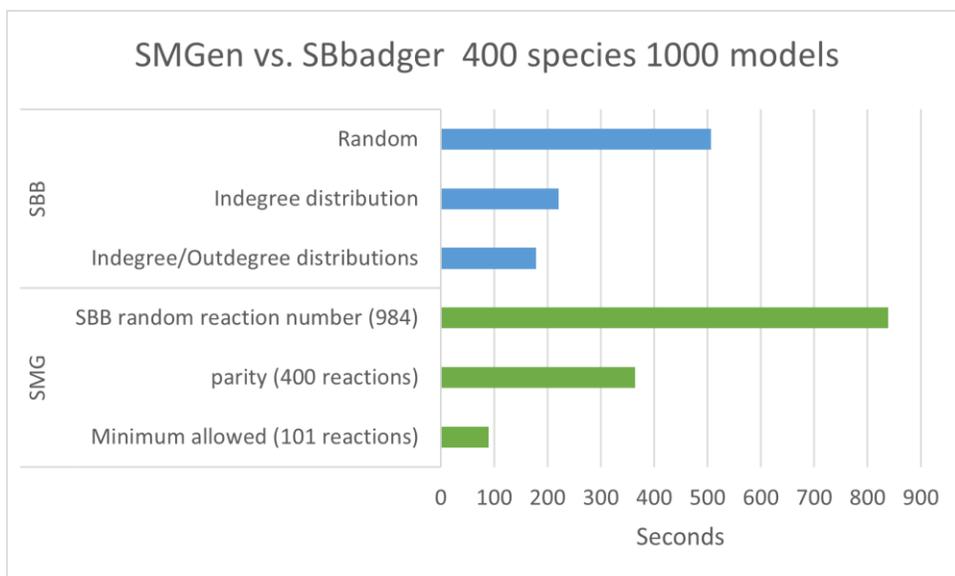

**Figure 7.** Runtime comparison for Sbbadger (SBB) and SMGen (SMG) when generating 1000 models with 400 species each. All runs were parallelized using all 8 cores of a Ryzen-7 4700U processor. The SBbadger runs are the random, indegree, and indegree/outdegree runs from Figure 6A. While the number of reactions in SBbadger generated models are distribution dependent, they are user-defined in SMGen. To obtain a reasonable comparison, SMGen was run with the number of reactions set to the average number of reactions for models in the SBbadger random run (984), parity with the number of species (400) and the minimum allowed (101).

| Major Differences | SBbadger | SMGen |
|---|---|---|
| Interface | Script/terminal | GUI |
| Modular workflow | Distributions, networks, and kinetics | NA |
| Degree distributions | Fully customizable indegree/outdegree distributions | Demonstrated to follow a power law distribution |
| Kinetics formalisms | Six types (see text) | Mass action |
| Reaction counts | Determined by distributions (minimum can be set if no distribution is defined) | Definable down to minimum needed for fully connected network |
| Reactant/Product counts | Max of 2 each | Fully customizable |
| Network consistency (mass balance) | Optional | NA |
| Parallelization | Yes | Yes |
| Output | Models, distribution and network information and optional figures. | Models |
| Model formats | SBML, Antimony | SBML, BioSimWare |

**Table 1.** Major feature differences between SBbadger and SMGen.

## Conclusions

Benchmarking is a pivotal step in the development, comparison, and differentiation of methods and software tools for the inference and simulation of biochemical reaction networks. Credible benchmarking relies on models that resemble those that tools were designed to process. SBbadger is designed to construct reaction or metabolic models with user-defined degree distributions that match the distributions of those networks targeted for analysis. In addition, six types of rate laws can be imparted on the reactions and many other options are available for additional flexibility. Three standard network types are also available, linear, cyclic, and branched. Models are output in both Antimony and SBML formats (Smith et al., 2009; Hucka et al., 2003).

## Acknowledgments

This work was supported by the National Cancer Institute under grant number U01CA242992. The content is solely the responsibility of the authors and does not necessarily represent the official views of the National Institutes of Health, the University of Washington or PNNL.

# Supplementary Information

SBbadger: Biochemical Reaction Networks with Definable Degree Distributions

Michael A. Kochen, H. Steven Wiley, Song Feng, and Herbert M. Sauro

Michael A. Kochen
E-mail: kochenma@uw.edu

This file includes Figures S1-S10, Tables S1-S4, and Text S1.

|       | Degree |       |       |       |       |       |
|-------|--------|-------|-------|-------|-------|-------|
|       | 1      | 2     | 3     | 4     | 5     | 6     |
| Prob. | 1      |       |       |       |       |       |
| Freq. | 50     |       |       |       |       |       |
|       |        |       |       |       |       |       |
| Prob. | 0.8    | 0.2   |       |       |       |       |
| Freq. | 40     | 10    |       |       |       |       |
|       |        |       |       |       |       |       |
| Prob. | 0.735  | 0.184 | 0.082 |       |       |       |
| Freq. | 36.735 | 9.184 | 4.082 |       |       |       |
|       |        |       |       |       |       |       |
| Prob. | 0.702  | 0.176 | 0.078 | 0.044 |       |       |
| Freq. | 35.122 | 8.78  | 3.902 | 2.195 |       |       |
|       |        |       |       |       |       |       |
| Prob. | 0.683  | 0.171 | 0.076 | 0.043 | 0.027 |       |
| Freq. | 34.162 | 8.541 | 3.796 | 2.135 | 1.366 |       |
|       |        |       |       |       |       |       |
| Prob. | 0.671  | 0.168 | 0.075 | 0.042 | 0.027 | 0.019 |
| Freq. | 33.526 | 8.381 | 3.725 | 2.095 | 1.341 | 0.931 |

**Table S1.** Example probability and frequency distribution generation for the power law function $k^{-2}/zeta(2)$ with 50 nodes and a minimum expected node frequency of 1. No degree range is provided. At each iteration the algorithm adds a degree to the distribution and recomputes the probabilities and expected values for the node frequencies. As degrees are added both the probabilities and the frequencies spread out. At iteration six the expected number of nodes with a frequency of 6 is less than the expected frequency threshold of 1 (red). The algorithm then stops and the probability mass function (PMF) from iteration five is selected.

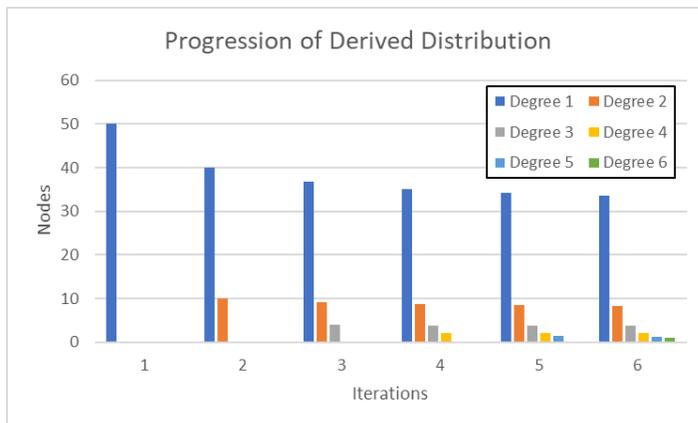

**Figure S1.** Progression of the frequency distribution derivation for the example in Table S1. At each iteration another degree bin is added, the probabilities are recalculated, and the nodes are redistributed. This repeats until the lowest expected node frequency is below the stated threshold. The broadest PMF with all degree bins above the threshold is accepted.

|  | Degree | | | | | | | |
|---|---|---|---|---|---|---|---|---|
|  | 1 | 2 | 3 | 4 | 5 | 6 | 7 | 8 |
| **Prob.** | 0.655 | 0.164 | 0.073 | 0.041 | 0.026 | 0.018 | 0.013 | 0.01 |
| **Freq.** | 32.735 | 8.184 | 3.637 | 2.046 | 1.309 | 0.909 | 0.668 | 0.511 |
|  |  |  |  |  |  |  |  |  |
| **Prob.** | 0.661 | 0.165 | 0.073 | 0.041 | 0.026 | 0.018 | 0.013 |  |
| **Freq.** | 33.073 | 8.268 | 3.675 | 2.067 | 1.323 | 0.919 | 0.675 |  |
|  |  |  |  |  |  |  |  |  |
| **Prob.** | 0.671 | 0.168 | 0.075 | 0.042 | 0.027 | 0.019 |  |  |
| **Freq.** | 33.526 | 8.381 | 3.725 | 2.095 | 1.341 | 0.931 |  |  |
|  |  |  |  |  |  |  |  |  |
| **Prob.** | 0.683 | 0.171 | 0.076 | 0.043 | 0.027 |  |  |  |
| **Freq.** | 34.162 | 8.541 | 3.796 | 2.135 | 1.366 |  |  |  |

**Table S2.** Example distribution generation for the power law function $k^2/zeta(2)$ with 50 nodes and a minimum expected node frequency of 1 node. A degree range of [1, 8] is provided. At each iteration the algorithm recomputes the probabilities and frequencies for each degree. If any have an expected frequency lower than the threshold, the lowest probability degree is removed (red), and another iteration begins. Once all expected frequencies are above the threshold the algorithm terminates.

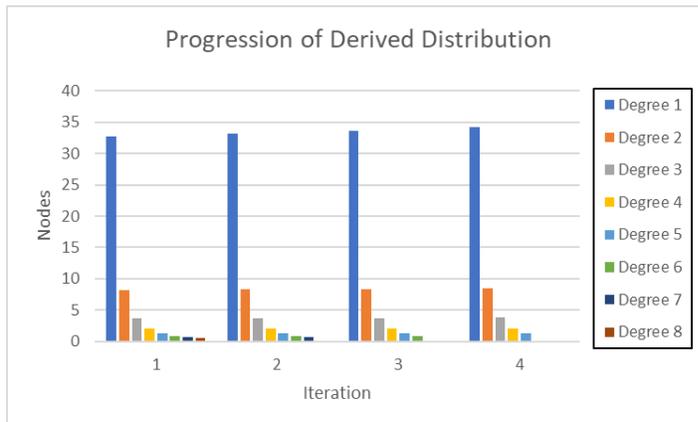

**Figure S2**. Progression of the frequency distribution derivation for the example in Table S2. At each iteration the degree with the lowest probability is removed and the probabilities and frequencies are recalculated. This repeats until all expected node frequencies are above the stated threshold. The final PMF with all degree bins above the threshold is accepted.

| Out Edges | 120.885 | | | | | | | |
|---|---|---|---|---|---|---|---|---|
| In Edges | 177.938 | | | | | | | |
| | | | | | | | | |
| Nodes | Expected Edges | | Degree | | | | | | |
| | | | 1 | 2 | 3 | 4 | 5 | 6 | 7 | 8 |
| 100 | 177.938 | Prob. | 0.655 | 0.164 | 0.073 | 0.041 | 0.026 | 0.018 | 0.013 | 0.01 |
| | | Freq. | 65.47 | 16.367 | 7.274 | 4.092 | 2.619 | 1.819 | 1.336 | 1.023 |
| | | | | | | | | | | |
| 99 | 176.158 | Prob. | 0.655 | 0.164 | 0.073 | 0.041 | 0.026 | 0.018 | 0.013 | 0.01 |
| | | Freq. | 64.815 | 16.204 | 7.202 | 4.051 | 2.593 | 1.8 | 1.323 | 1.013 |
| | | | | | | | | | | |
| 98 | 174.379 | Prob. | 0.655 | 0.164 | 0.073 | 0.041 | 0.026 | 0.018 | 0.013 | 0.01 |
| | | Freq. | 64.16 | 16.04 | 7.129 | 4.01 | 2.566 | 1.782 | 1.309 | 1.003 |
| | | | | | | | | | | |
| 97 | 166.363 | Prob. | 0.661 | 0.165 | 0.073 | 0.041 | 0.026 | 0.018 | 0.013 | |
| | | Freq. | 64.162 | 16.041 | 7.129 | 4.01 | 2.566 | 1.782 | 1.309 | |
| | | | | | | | | | | |
| ⋮ | | | | | ⋮ | | | | | ⋮ |
| | | | | | | | | | | |
| 75 | 128.63121 | Prob. | 0.661 | 0.165 | 0.073 | 0.041 | 0.026 | 0.018 | 0.013 | |
| | | Freq. | 49.61 | 12.402 | 5.512 | 3.101 | 1.984 | 1.378 | 1.012 | |
| | | | | | | | | | | |
| 74 | 121.565 | Prob. | 0.671 | 0.168 | 0.075 | 0.042 | 0.027 | 0.019 | | |
| | | Freq. | 49.618 | 12.405 | 5.513 | 3.101 | 1.985 | 1.378 | | |
| | | | | | | | | | | |
| 73 | 119.922 | Prob. | 0.671 | 0.168 | 0.075 | 0.042 | 0.027 | 0.019 | | |
| | | Freq. | 48.948 | 12.237 | 5.439 | 3.059 | 1.958 | 1.36 | | |

**Table S3.** Example distribution trimming. Once the outdegree and indegree distributions are sampled the total number of out-edges and in-edges must be equal. To make this more likely, the distribution with the higher number of edges, the indegree distribution in this example, is pared down. This is done by iteratively reducing the number of nodes and recalculating the distribution while adhering to the node-degree threshold. This produces a reduced expected value for, in this case, the in-edges. The distribution that exhibits the closest match between the in-edge and out-edge expected values is then chosen. The untouched distribution is sampled first, up to the number of desired species. The adjusted distribution is sampled up to the number of edges obtained when the first distribution is sampled. Multiple attempts to match the edge counts may be necessary and the algorithm will try `n_species` number of times.

| | Runtime Ratio (serial/parallel) | | |
|---|---|---|---|
| Num. Species | Random | Indegree | Indegree/Outdegree |
| 100 | 5.529682918 | 5.238575764 | 5.221457569 |
| 200 | 5.183895536 | 5.331865637 | 5.581884859 |
| 400 | 5.18849632 | 5.348644363 | 5.252087844 |
| 800 | 5.219227205 | 5.307022064 | 5.047741755 |

**Table S4.** Ratio of runtimes (serial/parallel) for the model generation times from Figures S10 and 6.

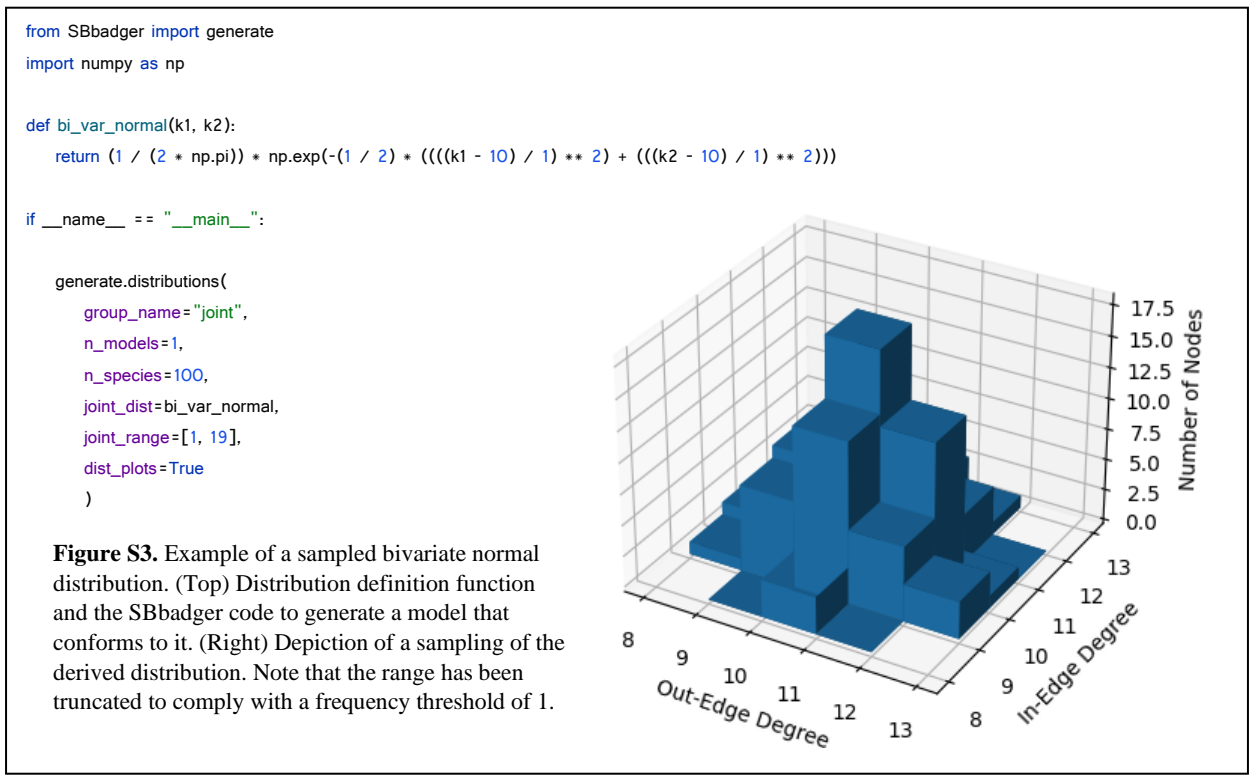

```python
from SBbadger import generate
import numpy as np

def bi_var_normal(k1, k2):
    return (1 / (2 * np.pi)) * np.exp(-(1 / 2) * ((((k1 - 10) / 1) ** 2) + (((k2 - 10) / 1) ** 2)))

if __name__ == "__main__":

    generate.distributions(
        group_name="joint",
        n_models=1,
        n_species=100,
        joint_dist=bi_var_normal,
        joint_range=[1, 19],
        dist_plots=True
    )
```

**Figure S3.** Example of a sampled bivariate normal distribution. (Top) Distribution definition function and the SBbadger code to generate a model that conforms to it. (Right) Depiction of a sampling of the derived distribution. Note that the range has been truncated to comply with a frequency threshold of 1.

**Text S1. Degree lists.** If a discrete list of probabilities is provided instead of a function, then the truncation or expansion algorithms in Tables S1 and S2 are skipped. These take the form of a list of tuples where the first term in each tuple is the deg and the second is the probability:

$$[(deg_1, prob_1), (deg_2, prob_2), \ldots, (deg_n, prob_n)].$$

If both distributions are provided, the edge adjustment algorithm, as seen in table S3, must still proceed. Lists of frequencies can also be provided. It that case both the probability and edge adjustments are skipped but in-edges and out-edges are checked for equality and an error is thrown if they are not. Frequency lists take the form:

$$[(deg_1, freq_1), (deg_2, freq_2), \ldots, (deg_n, freq_n)].$$

```python
from SBbadger import generate
from scipy.special import zeta

def in_dist(k):
    return k ** (-2) / zeta(2)

def out_dist(k):
    return k ** (-2) / zeta(2)

if __name__ == "__main__":

    model = generate.models(

        group_name='mass_action4',
        n_models=1,
        n_species=20,
        out_dist=out_dist,
        in_dist=in_dist,
        rxn_prob=[.35, .30, .30, .05],
        kinetics=['mass_action', ['loguniform', 'loguniform', 'loguniform',
                                  'loguniform', 'loguniform', 'loguniform',
                                  'loguniform', 'loguniform', 'loguniform',
                                  'loguniform', 'loguniform', 'loguniform'],
                  ['kf0', 'kr0', 'kc0',
                   'kf1', 'kr1', 'kc1',
                   'kf2', 'kr2', 'kc2',
                   'kf3', 'kr3', 'kc3'],
                  [[0.01, 100], [0.01, 100], [0.01, 100],
                   [0.01, 100], [0.01, 100], [0.01, 100],
                   [0.01, 100], [0.01, 100], [0.01, 100],
                   [0.01, 100], [0.01, 100], [0.01, 100]]],
        overwrite=True,
        rev_prob=.5,
        ic_params=['uniform', 0, 10],
        dist_plots=True,
        net_plots=True
    )
```

**Figure S4.** Example of mass action kinetics with parameter ranges defined for each parameter in each reaction pattern.

```python
from SBbadger import generate
from scipy.special import zeta

def in_dist(k):
    return k ** (-2) / zeta(2)

def out_dist(k):
    return k ** (-2) / zeta(2)

if __name__ == "__main__":

    model = generate.models(

        group_name='gma',
        n_models=1,
        n_species=10,
        out_dist=out_dist,
        in_dist=in_dist,
        rxn_prob=[.35, .30, .30, .05],
        kinetics=['gma', ['loguniform', 'loguniform', 'loguniform', 'uniform', 'uniform'],
                    ['kf', 'kr', 'kc', 'ko', 'kor'],
                        [[0.01, 100], [0.01, 100], [0.01, 100], [0, 1], [0, 1]]],
        gma_reg=[[0.5, 0.5, 0, 0], 0.5],
        overwrite=True,
        rev_prob=.5,
        ic_params=['uniform', 0, 10],
        dist_plots=True,
        net_plots=True
    )
```

**Figure S5.** Example of generalized mass action kinetics.

```python
From SBbadger import generate
from scipy.stats import zipf

def in_dist(k):
    return k ** (-2)

def out_dist(k):
    return zipf.pmf(k, 3)

if __name__ == "__main__":

    model = generate.models(

        group_name='lin_log',
        n_models=1,
        n_species=10,
        out_dist=out_dist,
        in_dist=in_dist,
        rxn_prob=[.35, .30, .30, .05],
        kinetics=['lin_log', ['uniform', 'uniform'],
                            ['v', 'rc'],
                            [[0.0, 100], [0.0, 100]]],
        overwrite=True,
        rev_prob=.5,
        ic_params=['uniform', 0, 10],
        dist_plots=True,
        net_plots=True
    )
```

**Figure S6.** Example of lin-log kinetics.

```python
From SBbadger import generate
from scipy.special import zeta

def in_dist(k):
    return k ** (-2) / zeta(2)

def out_dist(k):
    return k ** (-2) / zeta(2)

if __name__ == "__main__":

    model = generate.models(

        group_name='hanekom',
        n_models=1,
        n_species=10,
        out_dist=out_dist,
        in_dist=in_dist,
        rxn_prob=[.35, .30, .30, .05],
        kinetics=['hanekom', ['loguniform', 'uniform', 'loguniform'],
                              ['v', 'k', 'keq'],
                              [[0.01, 100], [0.0, 10], [0.01, 100]]],
        overwrite=True,
        rev_prob=.5,
        ic_params=['uniform', 0, 10],
        dist_plots=True,
        net_plots=True

    )
```

**Figure S7.** Example of generalized Michaelis-Menten (Hanekom) kinetics.

```python
From SBbadger import generate
from scipy.special import zeta

def in_dist(k):
    return k ** (-2) / zeta(2)

def out_dist(k):
    return k ** (-2) / zeta(2)

if __name__ == "__main__":

    model = generate.models(

        group_name='saturating_cooperative',
        n_models=1,
        n_species=20,
        out_dist=out_dist,
        in_dist=in_dist,
        rxn_prob=[.35, .30, .30, .05],
        kinetics=['saturating_cooperative', ['loguniform', 'loguniform', 'uniform', 'uniform'],
                                            ['v', 'k', 'n', 'nr'],
                                            [[0.01, 100], [0.01, 100], [0, 1], [0, 1]]],
        sc_reg=[[0.5, 0.5, 0, 0], 0.5],
        overwrite=True,
        rev_prob=.5,
        ic_params=['uniform', 0, 10],
        dist_plots=True,
        net_plots=True
    )
```

**Figure S8.** Example of saturable and cooperative kinetics.

```python
from SBbadger import generate
from scipy.special import zeta

def in_dist(k):
    return k ** (-2) / zeta(2)

def out_dist(k):
    return k ** (-2) / zeta(2)

if __name__ == "__main__":

    model = generate.models(

        group_name='modular_CM',
        n_models=1,
        n_species=10,
        out_dist=out_dist,
        in_dist=in_dist,
        rxn_prob=[.35, .30, .30, .05],
        kinetics=['modular_CM', ['loguniform', 'loguniform', 'loguniform', 'loguniform', 'loguniform',
                                 'loguniform', 'loguniform', 'loguniform', 'loguniform'],
                                ['ro', 'kf', 'kr', 'km', 'm',
                                 'kms', 'ms', 'kma', 'ma'],
                                [[0.01, 100], [0.01, 100], [0.01, 100], [0.01, 100], [0.01, 100],
                                 [0.01, 100], [0.01, 100], [0.01, 100], [0.01, 100]]],
        mod_reg=[[.5, .5, 0, 0], 0, .5],
        overwrite=True,
        rev_prob=0,
        ic_params=['uniform', 0, 10],
        dist_plots=True,
        net_plots=True
    )
```

**Figure S9.** Example of modular CM kinetics. The other four modular rate laws are `modular_DM`, `modular_SM`, `modular_FM`, and `modular_PM`. To obtain them simply insert them into the first position of the `kinetics` argument. Parameters are consistent among all five rate laws.

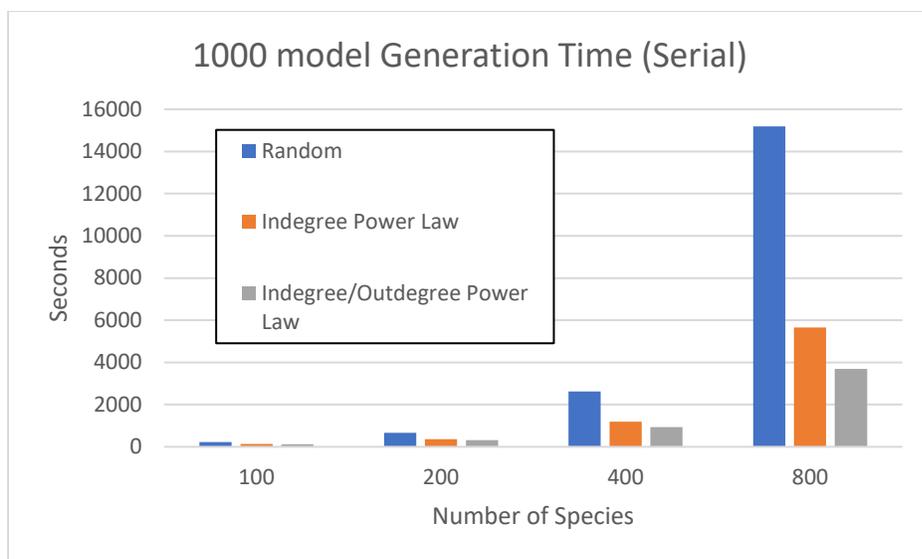

**Figure S10.** Time to completion for the serial creation of 1000 models with 100, 200, 400, and 800 species and three different degree distribution setups. One and two distribution cases are compared to the random selection of reaction species. For the indegree power law case the exponent is 2 and for the indegree/outdegree power law case the indegree exponent is 2 while the outdegree exponent is 3.